\documentclass[sigconf]{acmart}
\AtBeginDocument{%
  \providecommand\BibTeX{{%
    \normalfont B\kern-0.5em{\scshape i\kern-0.25em b}\kern-0.8em\TeX}}}

\setcopyright{acmcopyright}
\copyrightyear{2018}
\acmYear{2018}
\acmDOI{XXXXXXX.XXXXXXX}

\acmConference[Conference acronym 'XX]{Make sure to enter the correct
  conference title from your rights confirmation emai}{June 03--05,
  2018}{Woodstock, NY}
\acmPrice{15.00}
\acmISBN{978-1-4503-XXXX-X/18/06}

\usepackage{booktabs}   
\usepackage{subcaption} 

\usepackage{listings}
\usepackage{enumerate}
\usepackage{todonotes}
\usepackage{graphicx}
\usepackage{caption}
\usepackage{fancyvrb}
\usepackage{placeins}

\usepackage{multirow}
\usepackage{makecell}

\begin{document}

\title{Energy and Time Complexity for Sorting Algorithms in Java}


\author{Kristina Carter}
\email{kcarter@ruc.dk}
\affiliation{%
  \institution{Roskilde University}
    \country{Denmark}
  }

\author{Su Mei Gwen Ho}
\email{smgho@ruc.dk}
\affiliation{%
  \institution{Roskilde University}
    \country{Denmark}
}

\author{Mathias Marquar Arhipenko Larsen}
\email{mathl@itu.dk}
\affiliation{%
  \institution{IT University of Copenhagen}  
  \country{Denmark}
}

\author{Martin Sundman}
\email{sund@itu.dk}
\affiliation{
    \institution{IT University of Copenhagen}
      \country{Denmark}
}

\author{Maja H.\ Kirkeby}
\authornote{Corresponding author}
\email{kirkebym@acm.org}
\affiliation{%
  \institution{Roskilde University}
  \streetaddress{Universitetsvej 1}
  \country{Denmark}
  \postcode{4000}
  }
\renewcommand{\shortauthors}{}

\newcommand{\bigo}{$\mathcal{O}$} 
\newcommand{\bigomega}{$\Omega$} 
\begin{abstract}

\textbf{Background.} The article investigates the relationship between time complexity and energy consumption in sorting algorithms, focusing on commonly-used algorithms implemented in Java: Bubble Sort, Counting Sort, Merge Sort, and Quick Sort. The significance of understanding this relationship is driven by the increasing energy demands of Information and Communication Technology systems and the potential for software optimization to contribute to energy efficiency.

\textbf{Aims.} The main goal is to explore whether the theoretical time complexity of sorting algorithms can be used to predict energy consumption, by examining the correlation between time complexity and energy usage. If found, a strong correlation would enhance the ability of software developers to create energy-efficient applications.

\textbf{Method}. This quantitative study researches the execution of four selected sorting algorithms with input varying over input sizes (ranging from 25000 to 1 million) and input order types (best, worst, and random cases) on a single kernel in a Java-enabled system. The input size is adjusted according to the type's maximum execution time. A preliminary study found that the sufficient sample size varies from 30 to 400 measurements for each input and algorithm combination
. Thus, resulting in 136
combinations (not including the pre-study) totalling 12960 measurements. 
Wall time and the Central Processing Unit (CPU) energy consumption is measured; energy consumption is measured by CPU using Intel's Running Average Power Limit (RAPL) tool. Statistical analysis are used to examine the correlations between time complexity, wall time (execution time), and energy consumption.

\textbf{Results.} The study finds a strong correlation between time complexity and energy consumption for the sorting algorithms tested. More than 99\% of the variance in energy consumption for Counting Sort, Merge Sort, and Quick Sort depend on their time complexities. More than 94\% of the variance in energy consumption for Bubble Sort depends on its time complexity.

\textbf{Conclusions.}
The results affirm that time complexity can serve as a reliable predictor of energy consumption in sequential sorting algorithms. This discovery could guide software developers in choosing or designing algorithms that optimize energy efficiency. Future work will look into verifying these findings across different programming environments and hardware setups to generalize the concept of energy complexity.
\end{abstract}





\received{20 February 2007}
\received[revised]{12 March 2009}
\received[accepted]{5 June 2009}

\maketitle

\section{Introduction}

Information and Communication Technology (ICT) is an umbrella term used to describe all tools and systems which "collect, store, use and send data electronically"~\cite{Cambridge}. ICT includes server farms, computers, handheld devices such as smartphones, and more. It is estimated that 47\% of the world’s households have a computer today~\cite{itu}. Currently, these systems have contributed up to 10\% of the world’s energy consumption~\cite{burr}. As a conservative estimate, this represents approximately 3\% of the global carbon footprint~\cite{CO2}. The demands of ICT systems are continually increasing. The rate at which they are growing cannot be sustained and the infrastructure of these systems will collapse~\cite{ict}. As the world heavily relies on ICT systems, it is imperative to improve the energy consumption of such systems.

To date, there has been a focus on improving energy efficiency of ICT systems through hardware. However, it is believed that it is possible to achieve greater energy savings through the optimisation of the underlying software~\cite{Kumar2019,Field2014}. Despite this, there is a lack of understanding of energy consumption in software~\cite{Zecena2012}. 
The findings of~\cite{kumarEarly} indicate that software engineers do not program with energy efficiency in mind. This may stem from a tendency in computer science to construct programs with performance as the primary goal.

Sorting algorithms are essential in software and have been studied in relation to performance~\cite{kumarEarly}; time complexity expresses how input characteristics such as size and max value affect the performance of the algorithm. This abstract notion has been used by software developers to choose between different sorting algorithms, i.e., making an optimal choice when considering performance~\cite{introToAlgs}. Hence, it would be ideal to optimise the energy consumption of sorting algorithms if time complexity could be utilised as a guideline for energy consumption.

Multiple factors influence the energy consumption in IT systems~\cite{Li2017a} and while many of them are out of the control of the software developer, e.g., choice of hardware, number of concurrent users or internet quality, there are many that the software developer can influence directly, e.g., programming language~\cite{pereira2021}, data structures~\cite{Field2014,Kumar2019,Chandra2016} or choice of algorithms, e.g.,~\cite{Field2014,Chandra2016,Bunse2009}. While some studies on choice of algorithms have focused on energy consumption when executed on specific hardware, e.g., ~\cite{Schmitt2021,kirkeby2022energy,Bunse2009,jmaa2021}, this study focuses on the more abstract notion of time complexity when choosing algorithms. 
Previous studies disagree on whether time can be used as a guideline for energy consumption of sequential software; some studies indicate this~\cite{Mahmud2022,Lima2016}, whilst others show that execution time and energy consumption are not necessarily directly correlated~\cite{Field2014,Bunse2009}. 


In this work, we address the research question:
\begin{quote}
    \emph{How do time complexity and energy consumption relate for typical implementations of classic sorting algorithms?}
\end{quote}
Finding a correlation could improve existing state-of-the-art best practices for developing energy-efficient software, allowing developers to leverage existing research on improving algorithms to obtain better time complexities. 

We study four well-known sorting algorithms in the experiment: Bubble Sort, Counting Sort, Merge Sort and Quick Sort. These algorithms were chosen due to their high diversity in best and worst case time complexities, see Table \ref{tab:big-Oh}. They are all sequential in nature and we will investigate their  time complexity when executed in a sequential environment.
\begin{table}[]
    \centering
    \begin{tabular}{|c|c|c|}
         \hline
         Algorithm & Worst Case & Best Case \\
         \hline
         Bubble Sort & $O(n^2)$ & $\Omega(n^2)$ \\
         \hline
         Quick Sort & $O(n^2)$ &$\Omega(n log(n))$\\
         \hline
         Merge Sort & $O(n log(n))$ &$\Omega(n log(n))$\\
         \hline
         Counting Sort & $O(n+k)$ &$\Omega(n+k)$\\
         \hline  
    \end{tabular}
    \caption{Time complexity of algorithms studied~\cite{introToAlgs}. $n$ denotes input size, $k$ denotes the greatest value in the input.}
    \label{tab:big-Oh}
\end{table}
The sequential environment is ensured by implementing the sorting algorithms in a high-level sequential language and executing them on single kernels. In this study we have chosen Java, a highly popular language~\cite{IEEE_Top_Languages_2023,StackOverflow_2023_Survey,Statista_2023_Most_Used_Languages}
 with a suitable low variance in energy consumption~\cite{pereira2021}. 
 
While time complexity can be analysed in the high-level code, energy cannot; energy consumption depends on the hardware's execution of the algorithm. This is also the case for execution time; we cannot predict the wall time by only looking at high-level code. As seen in Equation~\ref{eq: energy}, energy and wall time are interdependent. Thus, this study is an experimental study specifically designed taking into consideration all factors which affect energy consumption; these occur in the sorting algorithms, the programming language, the implementations and I/O accesses, the measurement methodology, and the hardware. 


\paragraph{Contributions}
This work provides evidence that energy consumption is highly correlated with time complexity; for Counting Sort, Merge Sort and Quick Sort we show that more than 99\% of the variance is explained by the linear relationship between the energy consumption and the time complexity. For Bubble Sort it is more than 94\%.
In addition, we provide traceable evidence by calculating the correlation between time complexity and wall time, and between wall time and energy consumption.  

\paragraph{Article Structure}
In Section~\ref{sec:background} we provide basic theory on energy consumption and describe the four sorting algorithm implementations. Afterwards, we describe the experimental setup (Section~\ref{methodology}) and the results (Section~\ref{sec:results}). Lastly, we discuss the validity of the experiment (Section~\ref{sec:validity}), and compare the study to closely related work (Section~\ref{sec:related}).

\section{Background}\label{sec:background}
In the following we describe the basics of energy consumption and the implementations of the four sorting algorithms: Bubble Sort, Counting Sort, Merge Sort and Quick Sort.

\subsection{Energy}

To calculate the energy consumption of ICT systems, the energy definition of the ability to do work is used. This is written in the form:
\begin{equation}
     E_{energy} \equiv P_{ower} \times T_{ime}
     \label{eq: energy}
\end{equation}
where energy is measured in Joules (J). Power, the rate of energy consumption, is measured in Watts (W); 1 W is equal to 1 J of work done per second ~\cite{kumarEarly}.

In ICT systems, time in Equation \ref{eq: energy} refers to wall time, which includes central processing unit (CPU) processing time, memory access and all other processes needed during execution. Whilst time complexity describes the number of operations required in order to finish the program, wall time is dependant on CPU frequency~\cite{hardwarebook}.


\subsection{Sorting Algorithms}

Sorting algorithms are a fundamental concept of computer science~\cite{introToAlgs}. These algorithms take inputs such as lists, arrays, etc. and rearrange the elements in either ascending or descending order. Each sorting algorithm is designed and optimised for various kinds of input. In the following section, the sorting algorithms used in this paper are explained in further detail. 

\subsubsection{Bubble Sort}

\begin{figure}
\begin{lstlisting}[frame=lines,
basicstyle=\footnotesize, %or \tiny, \small \footnotesize etc.
language=C
]
BubbleSort (A)
    for i = 0 to A.length - 1
        for j = 0 A.length - 2 - i
            if A[j] > A[j + 1] then 
                swap (A[J], A[J + 1])
\end{lstlisting}
\caption{Bubble Sort pseudocode \cite{dataStructAlgs}.}
\label{fig:bubblePseudo}
\end{figure}
Bubble Sort is a simple and popular sorting algorithm~\cite{introToAlgs, bubbleReview}. It organizes arrays by comparing two elements at a time and swapping them if they are not in the right order. This action is repeated until the algorithm has iterated through the entire array. Bubble Sort completes this iteration by using a nested-loop detailed in Figure~\ref{fig:bubblePseudo}, hence its time complexity of $n^2$.

\subsubsection{Counting Sort} 

\begin{figure}
\begin{lstlisting}[frame=lines,
basicstyle=\footnotesize, %or \tiny, \small \footnotesize etc.
language=C
]
CountingSort(A, B, k)
    // let C[0..k] be a new array 
    for i = 0 to k
        C[i] = 0
    for j = 1 to A.length
        C[A[j]] = C[A[j]] + 1
    /* C[i] now contains the number of elements 
        equal to i */
    for i = 1 to k
        C[i] = C[i] + C[i - 1]
    /* C[i] now contains the number of elements 
        less than or equal to i */
    for j = A.length down to 1
        B[C[A[j]]] = A[j]
        C[A[j]] = C[A[j]] - 1
\end{lstlisting}
\caption{Counting Sort pseudocode \cite{introToAlgs}.}
\label{fig:countingPseudo}
\end{figure}

Counting Sort does not utilise comparison for sorting. Instead, it "counts" the frequency of occurrences of each element in the array \cite{introToAlgs}. By accessing only the array, Counting Sort has a linear time complexity. However, Counting Sort uses temporary arrays as seen in Figure~\ref{fig:countingPseudo}, requiring relatively more memory than some other sorting algorithms.

\subsubsection{Merge Sort}

\begin{figure}
\begin{lstlisting}[frame=lines,
basicstyle=\footnotesize, %or \tiny, \small \footnotesize etc.
language=C
]
MergeSort(A, p, r) //divide into subarrays
    if p < r
        q = (p + r) / 2
        MergeSort(A, p, q)
        MergeSort(A, q + 1, r)
        Merge(A, p, q, r)

Merge(A, p, q, r) //merge subarrays
    n_1 = q - p +1
    n_2 = r - q
    // let L[1..n_1+1] and R[1..n_2+1] be new arrays
    for i = 1 to n_1
        L[i] = A[p + i - 1]
    for j = 1 to n_2
        R[j] = A[q + j]
    L[n_1 + 1] = max
    R[n_2 + 1] = max
    i = 1
    j = 1
    while i < n1 + 1 and j < n2 + 1
        if L[i] =< R[j]
            A[k] = L[i]
            i = i + 1
        else
            A[k] = R[j]
            j = j + 1

    while i < n1 + 1
        A[k] = L[i]
        i = i + 1
        k = k + 1

    while j < n2 + 1
        A[k] = R[j]
\end{lstlisting}
\caption{Merge Sort pseudocode \cite{introToAlgs}.}
\label{fig:mergePseudo}
\end{figure}

Merge Sort is a comparison sorting algorithm~\cite{Bunse2009}. It uses a divide-and-conquer strategy, separating its input array into smaller and smaller arrays until each array consists of only one element. Once the original array is completely split, the algorithm then sorts and merges recursively ~\cite{introToAlgs}, see Figure \ref{fig:mergePseudo}. Its time complexity is $n\log(n)$.

\subsubsection{Quick Sort}

\begin{figure}
\begin{lstlisting}[frame=lines,
basicstyle=\footnotesize, %or \tiny, \small \footnotesize etc.
language=C
]
QuickSort(A as Array, p as int, r as int)
    if {p < r}
        q = Partition(A, p, r)
        QuickSort(A, p, q - 1)
        QuickSort(A, q + 1, r)

Partition(A as array, p as int, r as int)
    pivot = A[r]
    i = p - 1
    for j = p to r - 1
        if A[j] =< pivot
            i = i + 1
            swap A[i] with A[j] 
    swap A[i + 1] with A[r]
    return i + 1
\end{lstlisting}
\caption{Quick Sort pseudocode \cite{introToAlgs}.}
\label{fig:quickPseudo}
\end{figure}

Quick Sort is another divide-and-conquer algorithm. The key difference between these Merge Sort and Quick Sort is the pivot that Quick Sort implements~\cite{introToAlgs}, see Figure \ref{fig:quickPseudo}. Depending on the pivot chosen, the time complexity of Quick Sort can be $n\log(n)$ or $n^2$. 

\section{Methodology}\label{methodology}

Here, the physical setup of the experiment is detailed. Then, the input sizes and cases are listed. This leads to the method of gathering the data. Finally, the method implemented of analysing the results is explained. 

\subsection{Experimental Setup}

The experiment was originally carried out on four laptops, IDs 6, 16, 21, and 23, to check if each laptop produced similar results. After ensuring similar behaviour, we randomly chose to present data from laptop ID 16. The laptops used in the experiment are Lenovo ThinkPad x260, with Intel(R) Core(TM) i5-6200U CPU \footnote{see \url{https://ark.intel.com/content/www/us/en/ark/products/88193/intel-core-i56200u-processor-3m-cache-up-to-2-80-ghz.html}}, with max frequency 2.30GHz and a \verb|x86_64| architecture with 2 cores, 3MiB Cache, 8GB Memory, 256GB Storage. On them we installed Ubuntu 22.04 LTS and OpenJDK 11.0.19.

The four laptops were placed in a temperature-regulated glass box set to 20$^{\circ}$C. Here, a thermostat monitored the temperature and regulated an air conditioning unit. The laptops were inspected once a day, where the time, temperature, and status of the laptops was logged and controlled; the temperature was logged to always be within 19.6$^{\circ}$C and 20.4$^{\circ}$C. 

\subsection{Energy Estimation}

We employed the method presented in ~\cite{pereira2021} to estimate the energy consumption of the software, using Intel's RAPL to retrieve information on the energy consumption of the machine. This provided us with the required relative precision of the measurements; the energy consumption measured was not the accurate energy consumption, but the pair-wise relation between measurement of two executions was correct and useful.
A consequence is that the estimates may be skewed from the ground truth and the actual numbers cannot be used for comparison when executed on other machines. A preliminary studies showed that the CPU is the main energy consumer~\cite{BScThesis}; this aligns with the typical power consumption of CPU and dynamic random access memory (DRAM) ~\cite{Lim2008}. Therefore, for the purpose of data analysis, the CPU energy measurements are used instead of RAPL's Package or PKG energy measurements.
%

\begin{figure} 
\begin{lstlisting}[frame=lines,
basicstyle=\footnotesize, %or \tiny, \small \footnotesize etc.
language=C
]
for (i = 0; i < n; i++)
{
  // reads time before
  gettimeofday(&tvb, 0);  
  // using rapl.c, reads rapl reg.
  rapl_before(fp_csv, core); 

  // execute Java program
  system(command);   

  // use rapl.c calc. energy cons. (J)   
  rapl_after(fp_csv, core); 
  // read and calc time cons. (ms)
  gettimeofday(&tva, 0);    
  time_spent = ((tva.tv_sec - tvb.tv_sec) 
               *1000000 + tva.tv_usec - tvb.tv_usec) 
               / 1000;
}
\end{lstlisting}
\caption{Obtaining the time and energy spent on executing the Java program~\cite{pereira2021}.}
\end{figure}

\subsection{Experimental Implementations of Algorithms}
The algorithms used were introduced in the previous section; the Java implementations follow the structure of the Figures~\ref{fig:bubblePseudo}, \ref{fig:countingPseudo}, \ref{fig:mergePseudo}, and \ref{fig:quickPseudo}. The algorithms are called in a similar fashion, shown in Figure ~\ref{fig:Sort-Deepcopy}. Data structures are aligned and limited to arrays and integers.

\subsection{Time Complexity and Input Space}
%
In this section, the input sizes for each experiment are listed, and the type of each input case is detailed. 

After a series of preliminary tests, input sizes ranging from 25000 to 1 million, and intervals ranging from 25000 to 100000, were chosen to give clear, uncluttered results across the chosen algorithms and time complexities, see Table \ref{tab:TestSpace}. 
In particular, the input size was capped at 500000 for Bubble Sort and 200000 for Quick Sort worst case due to time constraints, bearing in mind that they have a worst case time complexity of O($n^2$), see Table \ref{tab:big-Oh}. 

The final set of input sizes for all tests was decided: 25000, 50000, 75000, 100000, 200000, 300000, 400000, 500000, 600000, 700000, 800000, 900000 and 1 million. Bubble Sort tests stopped at 500000. Only Quick Sort worst case input size intervals varied to the rest: 25000, 50000, 75000, 100000, 125000, 150000, 175000 and 200000; these smaller intervals allowed a closer view on the behaviour of the time complexity.

The inputs were manipulated in order to create three specific cases for each algorithm: best case, worst case, and what we will refer to as "random case". Table~\ref{tab:TestSpace} details the various input types assigned to the best and worst cases for each algorithm. Best and worst case input types were designed to reflect textbook definitions, i.e., the input that would trigger the best case time complexity and worst case time complexity, respectively. The random case using randomly-sorted inputs was devised as an extra check. Counting Sort was an exception without a random case, as the best and worse case inputs were already randomly-sorted. 


\begin{table*} \footnotesize
    \centering
    \begin{tabular}{l@{\;\;}l@{\;\;}l@{\;\;}l@{\;\;}l@{\;\;}l@{\;\;}l} \toprule
    Algorithm	&Case	&Time Complexity	&Input Type	&Runs	&Value Range	&Input Range\\ \midrule
Bubble Sort &Worst case	&\bigo($n^2$)	&Reverse-sorted	&1 CSV x 400	&0 - size	&25k - 500k\\
	   &Best case	&\bigomega($n^2$)	&Sorted	&1 CSV x 400	&0 - size	&25k - 500k\\
	&Random case	&-	&Randomly sorted	&10 CSVs x 40	&-max - +max	&25k - 500k\\\midrule
Merge Sort	&Worst case	&\bigo($n\log(n)$)	&Alternating Elements	&1 CSV x 30	&0 - size	&25k - 1 mil\\
	&Best case	&\bigomega($n\log(n)$)	&Sorted	&1 CSV x 30	&0 - size	&25k - 1 mil\\
	&Random case	&-	&Randomly sorted	&10 CSVs x 3	&-max - +max	&25k - 1 mil\\\midrule
Quick Sort	&Worst case	&\bigo($n^2$)	&Reverse-sorted	&1 CSV x 30	&0 - size	&25k - 200k\\
	&Best case	&\bigomega($n\log(n)$)	&Evenly-partitioned	&1 CSV x 30	&0 - size	&25k - 1 mil\\
	&Random case	&-	&Randomly sorted	&10 CSVs x 3	&-max - +max	&25k - 1 mil\\\midrule
Counting Sort	&Worst case	&\bigo($n + k$)	&Randomly sorted, big k	&1 CSV x 30	&0 - 100 mil	&25k - 1 mil\\
	&Best case	&\bigomega($n + k$)	&Randomly sorted, small k	&1 CSV x 30	&0 - 10	&25k - 1 mil\\\midrule
ReadCSV	&Baseline	&-	&Sorted	&1 CSV x 400	&0 - size	&25k - 1 mil\\
	&	&	&Reverse-sorted	&1 CSV x 400	&0 - size	&25k - 1 mil\\
	&	&	&Randomly sorted	&10 CSV x 40	&-max - +max	&25k - 1 mil\\
	&	&	&Alternating Elements	&1 CSV x 400	&0 - size	&25k - 1 mil\\
	&	&	&Evenly-partitioned	&1 CSV x 400	&0 - size	&25k - 1 mil\\
	&	&	&Counting worst*	&1 CSV x 400	&-max - +max	&25k - 1 mil\\
	&	&	&Randomly sorted, small k	&10 CSVs x 40	&0 - 10	&25k - 1 mil\\
	&	&	&Randomly sorted, big k	&10 CSVs x 40	&0 - 100 mil	&25k - 1 mil\\ \bottomrule
    \end{tabular}
    \caption{The test space for the study, comprising of four algorithms and one baseline. This table shows the time complexity for each case of each sorting algorithm, the input types used, the number of runs performed, the value ranges and the input ranges. Less runs were required in tests which used a  ``deep copy''-approach due to lower variance.} 
    \label{tab:TestSpace}
\end{table*}

\subsection{Preliminary Study: Importing and Sorting Inputs Ratio}
We conducted the preliminary study in March 2023 to May 2023. The sample sizes in the preliminary study were based on Cochrane's formula for infinite populations with 95\% confidence, i.e., requiring 385 samples for each combination of algorithm and input. 



In order to support sorting different inputs, each sorting algorithm first imported the specific input from a CSV file into an array, then sorted the array, see Figure~\ref{fig:Sort-CsvtoArray}.  This standard import of inputs required much time on I/O compared to how much time was spent on sorting afterwards.  
Thus, a fifth program called "ReadCSV" was tested as shown in Table \ref{tab:TestSpace}. It would provide a baseline for the sorting algorithm energy and time measurements, as it included importing the CSV file into an array, as well as any unmitigated background processes. The method would allow us to subtract the baseline measurements from the sorting algorithm data and provide clearer results. 

However, the import of the input from a CSV into an array was more time consuming than the sorting of the array, introducing noise to the results. This was reflected in the not-normal distribution in the harvested data, see Table ~\ref{tab:ND}, indicating the occurrence of uncontrolled variables~\cite{cruz2021green}. It should be noted that although noise may affect the statistical analysis, the methodology for checking correlation is still applicable for non-normal distributed data. Only for one algorithm, namely Bubble Sort, was the ratio between importing the input and sorting the input suitable; the time over input reflected the expected time complexity ~\cite{BScThesis}).

During this study, we needed to adjust the ratio between the time spent on importing input and the time sorting it in order to get viable data. We introduced a ``deep copy'' of the array and sorted the array $400$ times per import, see Figure~\ref{fig:Sort-Deepcopy}. The choice of $400$ repetitions was based on Bubble Sort's ratio between time spent on input import and time sorting obtained in the preliminary study ~\cite{BScThesis}. This solution has one caveat: repetitions within the same file may have caused the compiler to optimize across each call to the sorting algorithms; however, each algorithm had the same advantages/disadvantages across the entire input space. This "deep copy" method was used for all tests except for the ReadCSV tests.

\begin{figure}\footnotesize
\begin{subfigure}[t]{0.45\textwidth}
\begin{lstlisting}[frame=lines,
basicstyle=\footnotesize, %or \tiny, \small \footnotesize etc.
language=Java
]
public static void main(String[] args ){
  int[] arr =  CSV_toArr(args[0],args[1]);
  sort(arr);
}
\end{lstlisting}
\caption{}
\label{fig:Sort-CsvtoArray}
\end{subfigure}

\begin{subfigure}[t]{0.45\textwidth}
\begin{lstlisting}[frame=lines,
basicstyle=\footnotesize, %or \tiny, \small \footnotesize etc.
language=Java
]
public static void main(String[] args ){
  int[] arr =  CSV_toArr(args[0],args[1]);
  for (int i = 0; i < 400; i++){
    int[] arrCopy = new int[arr.length];
    System.arraycopy(arr,0,arrCopy,0,arr.length);
    sort(arrCopy);
  }
}
\end{lstlisting}
\caption{}
\label{fig:Sort-Deepcopy}
\end{subfigure}
\caption{As seen in Figure ~\ref{fig:Sort-CsvtoArray}, the inital method was to first import an input from a CSV file into an array, then sorting it. This resulted in the import of the input being more time consuming than sorting of it, introducing noise into the data, see Table ~\ref{tab:ND}. 
Figure ~\ref{fig:Sort-Deepcopy} shows how we adjusted the ratio between time spent on importing input and time spent sorting it; we introduced a deep copy of the array and sorted the array 400 times per import.}
\label{fig:Sort-CsvtoArray-Deepcopy}
\end{figure}


\subsection{Data Collection and Cleaning}
The data was collected after four months of active testing during two time periods: the preliminary study in March 2023 to May 2023 and "deep copy" study in June 2023 to August 2023.

As mentioned, the sample sizes in the preliminary study was based on Cochrane's formula, thus, we obtained 400 samples for each combination of algorithm and input. A test-run of the "deep copy" method had a smaller variance and the sample size was therefore reduced to 30 samples, see Table~\ref{tab:TestSpace}, column ``Runs''.

The energy consumption data from the Bubble Sort ``deep copy'' tests contained inviable data due to overflow. These overflows are well-known~\cite{RAPLinAction,pereira2021} and are a result of the design of the register on the Intel chip which is limited to 32 bits; the higher the wall time, the more likely the register will overflow \cite{RAPLinAction}. For the purposes of analysis, we decided to use the Bubble Sort data from the preliminary study along with the "deep copy" data from Merge, Counting and Quick Sort.

All negative energy values were removed\footnote{This procedure follows~\cite{pereira2021}.} from the harvested data, and 
the baseline subtracted, i.e., the time and energy used to import the CSV.
Afterwards, the outliers were removed.
The data for each input and algorithm combination was tested for normal distribution using the Shapiro-Wilks method.




\subsection{Statistical Methodology: Checking Non-Linear Correlation}\label{sec:corr-appr}

There are ways to perform a linear regression for $n+k$ and $n^2$, however, there are no standard ways to test for correlation with $n\log(n)$ growth. To have a singular approach to comparing wall time and energy consumption to time complexities, a simple solution was found: we converted the input size $n$ on the x-axis to that of the time complexity, allowing us to perform a linear regression and examine the correlation between the time (or energy) and time complexity. 

\begin{figure*}
    \centering
    \begin{subfigure}[b]{0.45\textwidth}
         \centering
         \includegraphics[width=\textwidth]{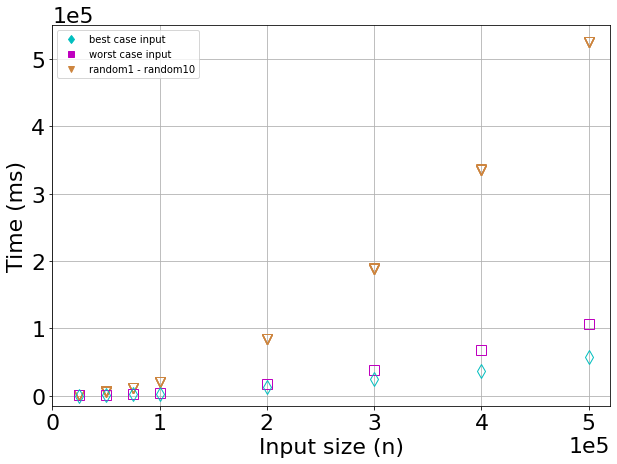}
         \caption{}
         \label{fig:bubble_WT_n-demo}
     \end{subfigure}
    \begin{subfigure}[b]{0.45\textwidth}
         \centering
         \includegraphics[width=\textwidth]{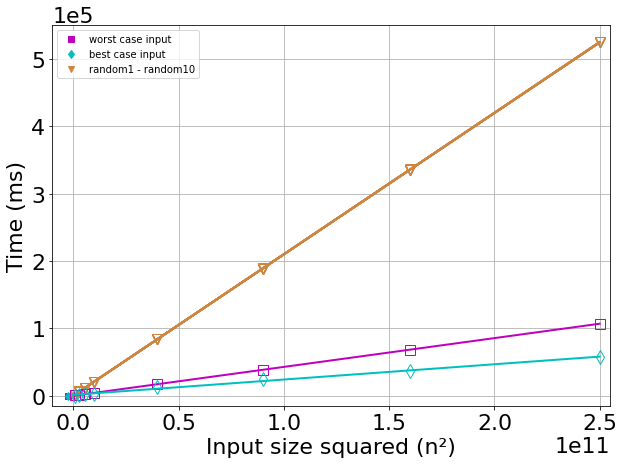}
         \caption{}
         \label{fig:bubble_WT_bigO-demo}
    \end{subfigure}    
    \caption{Example of how we check for correlation, where Bubble Sort input size $n$ is converted to time complexity $n^2$ on the x-axis, such that a correlation with wall time can be investigated. a) Bubble Sort wall time plotted against input size $n$. b) Bubble Sort wall time plotted against time complexity, represented as $n^2$. Note the scientific notation on both axes.}
    \label{fig:bubble_convert}
\end{figure*}

To demonstrate this approach, we use Bubble Sort's time complexity and wall time shown in  Figure~\ref{fig:bubble_WT_n-demo}. Here, we see the average wall time for each case (best, worst or random) plotted against the input size. 
Instead of considering the \emph{complexity-converted function} $bubblesort\colon \mathit{size}(\mathit{input}) \rightarrow \mathit{time}$ where $\mathit{size}$ returns the length of the input list $\mathit{input}$, we consider the function
$bubblesort'\colon$  $\mathit{size'}(\mathit{input}) \rightarrow \mathit{time}$, where $\mathit{size'}$ returns the square of the input size.
In other words, we convert x-axis values from $n$ to $n^2$, shown in Figure \ref{fig:bubble_WT_bigO-demo}, 
resulting in updated data points to which a regression line can be fitted. 

Once the regression line has been fitted, we calculate the coefficient of determination, $R^2$, which describes how much of the change in wall time in relation to the time complexity (in this case), directly depends on the time complexity. The $R^2$ indicates how well the linear regression fits the data, and therefore how closely the two variables are correlated.
\footnote{\label{r2-def}The coefficient of determination, $R^2$, describes the proportion of variance in the dependent variable (Y) which is explained by the linear relationship between the independent and dependent variables (X and Y). The $R^2$ value is always between 0 and 1, where the closer the regression line fits the data, the closer $R^2$ will be to 1 ~\cite{Samuels2025}.}

\section{Results}\label{sec:results}

Our primary result obtained in this study is a strong correlation between energy consumption and time complexity. Counting Sort, Merge Sort, Quick Sort have especially high $R^2$ values greater than 99\%, while Bubble Sort $R^2 \geq 0.94$, indicating that variance of energy consumption is closely linked to time complexity.


Our secondary finding is that the theoretical worst case input does not yield the practical worst case energy consumption nor wall time for both Bubble Sort and Merge Sort. Both algorithms show the random case to have slower performance and use more energy than the worst case. 


\subsection{Primary Finding: Energy Consumption and Time Complexity}\label{sec:RQ}
\begin{figure*}
    \centering
    \begin{subfigure}[b]{0.33\textwidth}
         \centering
         \includegraphics[width=\textwidth]{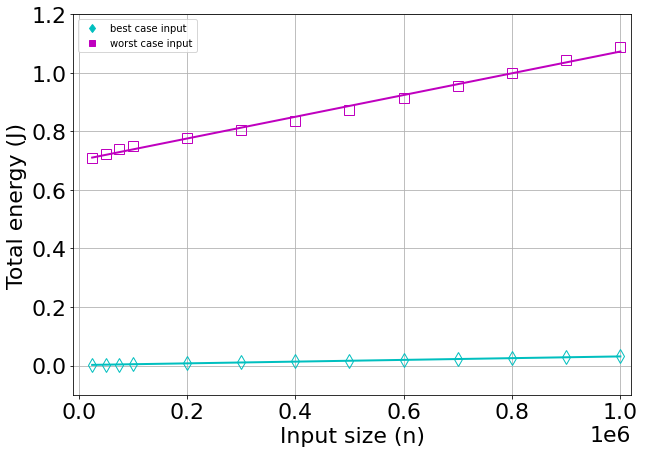}
         \caption{Counting Sort,  O(n+k)}
         \label{fig:counting-EC-bigO}
     \end{subfigure}
    \begin{subfigure}[b]{0.33\textwidth}
         \centering
         \includegraphics[width=\textwidth]{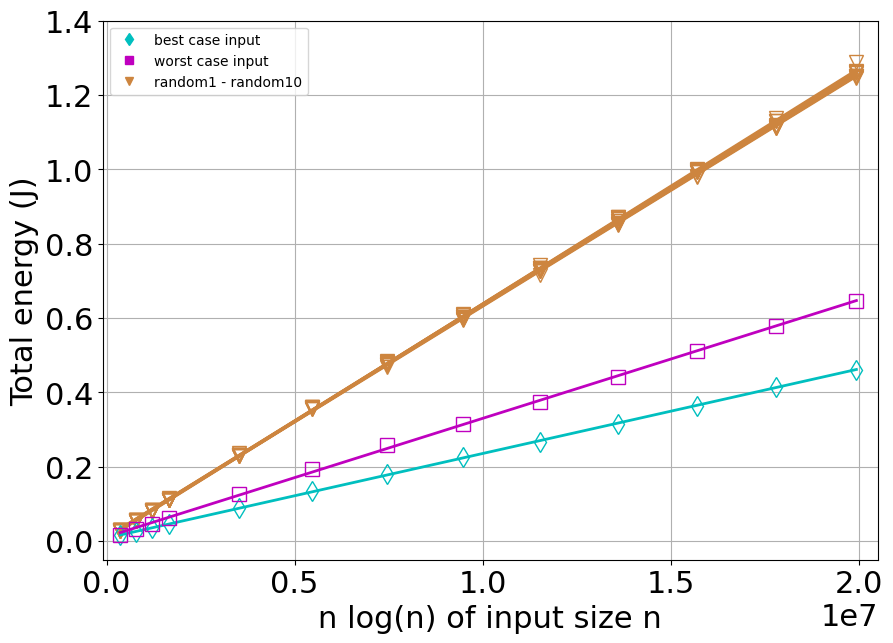}
         \caption{Merge Sort, O(n log(n))}
         \label{fig:merge_EC_bigO}
     \end{subfigure} 
     \begin{subfigure}[b]{0.33\textwidth}
         \centering
         \includegraphics[width=\textwidth]{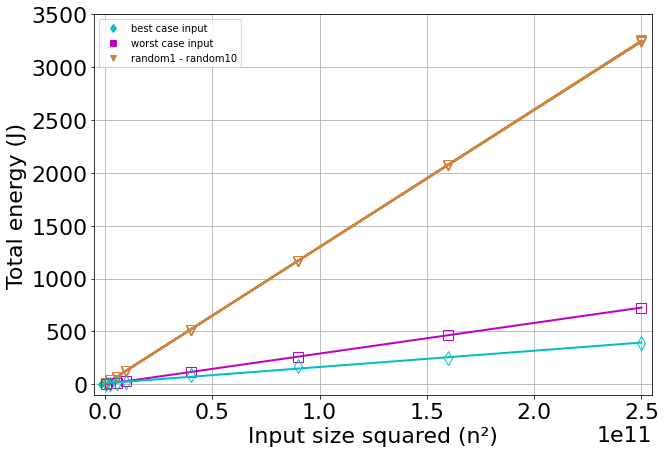}
         \caption{Bubble Sort, O($n^2$)}
         \label{fig:bubble_EC_bigO}
     \end{subfigure} 
     \begin{subfigure}[b]{0.33\textwidth}
         \centering
         \includegraphics[width=\textwidth]{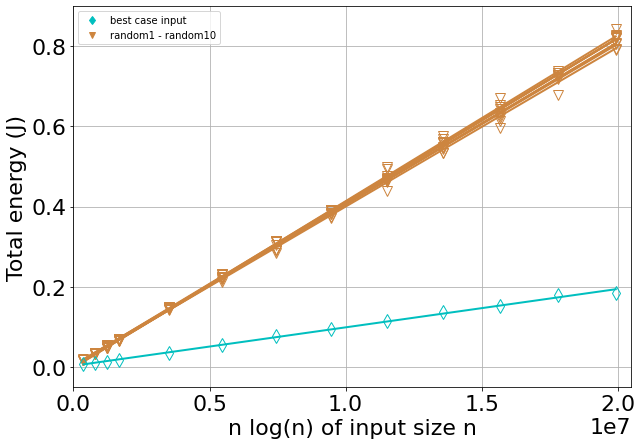}
         \caption{Quick Sort best and random case, O(n log(n))}
         \label{fig:quick1_EC_bigO}
     \end{subfigure}    
     \begin{subfigure}[b]{0.33\textwidth}
         \centering
         \includegraphics[width=\textwidth]{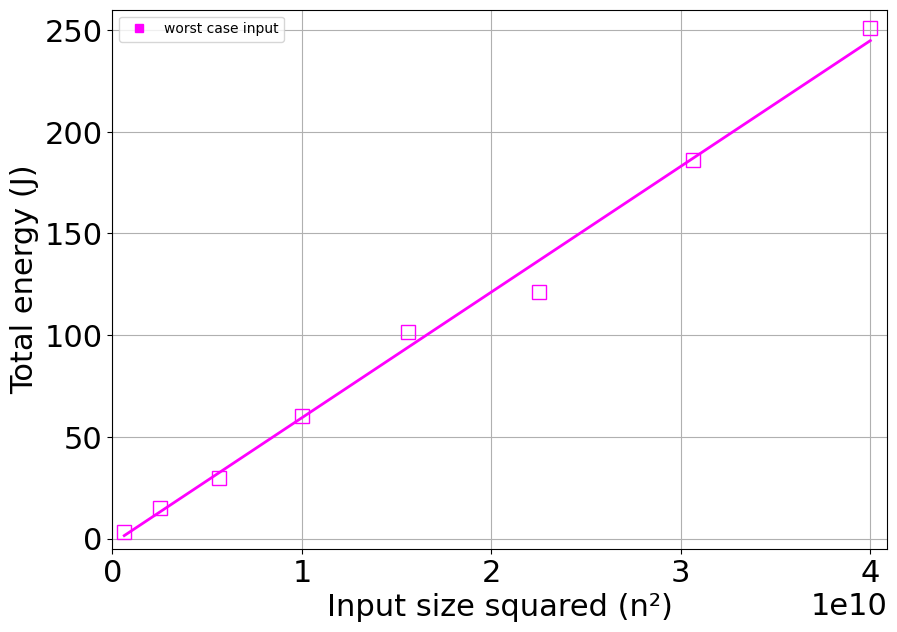}
         \caption{Quick Sort worst case, O($n^2$)}
         \label{fig:quick2_EC_bigO}
     \end{subfigure}    
        
    \caption{Energy consumption against time complexity, where the x-axis is adjusted according to the BigO of the sorting algorithm. Note the scientific notation on the x-axes.}
    \label{fig:All-EC-bigO}
\end{figure*}

Figure~\ref{fig:All-EC-bigO} shows the relation between energy consumption and time complexities for the test space; Bubble Sort~(Figure~\ref{fig:bubble_EC_bigO}), 
Counting Sort~(Figure~\ref{fig:counting-EC-bigO}), 
Merge Sort~(Figure~\ref{fig:merge_EC_bigO})
and Quick Sort~(Figures~\ref{fig:quick1_EC_bigO} and~\ref{fig:quick2_EC_bigO}). The inputs generated to create the theoretical best case are blue, the inputs generated to demonstrate the theoretical worst case are pink and the random cases are brown.

There is a clear trend of linear correlation between time complexity and energy consumption in all five figures. Note that for Counting Sort, the energy consumption is plotted against input size as time complexity is \bigo$(n+k)$; the larger $k$-value $10000000$ is reflected in the increased gradient of the worst case slope~(Figure~\ref{fig:counting-EC-bigO}, pink line).

Counting Sort best case~(Figure~\ref{fig:counting-EC-bigO}, blue line) is the most energy efficient for $n = 1000000$, followed by Quick Sort best case~(Figure~\ref{fig:quick1_EC_bigO}, blue line) and Merge Sort best case~(Figure~\ref{fig:merge_EC_bigO}, blue line).
The least energy-efficient sorting algorithms for $n = 500000$ are all time complexity \bigo($n^2$); Bubble Sort random case~(Figure~\ref{fig:bubble_EC_bigO}, brown lines), Quick Sort worst case~(Figure~\ref{fig:quick2_EC_bigO}, pink line) and Bubble Sort worst case~(Figure~\ref{fig:bubble_EC_bigO}, pink lines).
Merge Sort random case~(Figure~\ref{fig:merge_EC_bigO}, brown lines) shows a more stable/consistent energy use than Quick Sort random case~(Figure~\ref{fig:quick1_EC_bigO}, brown line). 

The obtained measurements provide evidence supporting a strong correlation between energy consumption and various time complexities, see Table~\ref{tab:EC-bigO-Rvalues}\footnote{\label{R-round}Rounded to 9 decimal places. Note that when the final digit is 9, the value is rounded to 8 decimal places.}\label{R-round}, i.e., $R^2 \geq 0.94$. 
For most of the algorithms, namely Counting Sort, Merge Sort and Quick Sort the coefficient of determination is highest, i.e., $R^2 \geq 0.99$ \ref{r2-def}. Bubble Sort has a slightly lower $R^2 \geq 0.94$; bear in mind that the the method for obtaining Bubble Sort measurements differed from the "deep copy" method used for the other algorithms due to overflow of the energy register. 
The high coefficient of determination for all four algorithms indicate that for this setup, time complexity can be used as a model for predicting energy consumption of sorting algorithms. 

\begin{table}[h]
\centering
\caption{The coefficient of determination ($R^2$) between energy consumption and time complexity for sorting algorithms $^{4}$.}
\begin{tabular}{@{}l@{\;}l@{\;}l@{\;}l@{\;}l@{}} \toprule
&Bubble Sort	&Counting Sort	&Merge Sort	&Quick Sort\\ \midrule
Best Case	&0.96980903	&0.999236567	&0.999923137	&0.997438796\\
Worst Case	&0.945110649	&0.994077848	&0.999490504	&0.993539115\\
Random Case	&	&	&	&\\
$\quad$random\_1	&0.944638742	&\multicolumn{1}{c}{-}	&0.999791571	&0.998894583\\
$\quad$random\_2	&0.944509971	&\multicolumn{1}{c}{-}	&0.999820907	&0.999071556\\
$\quad$random\_3	&0.944911875	&\multicolumn{1}{c}{-}	&0.999870928	&0.999294277\\
$\quad$random\_4	&0.944155982	&\multicolumn{1}{c}{-}	&0.999879961	&0.996371686\\
$\quad$random\_5	&0.944766841	&\multicolumn{1}{c}{-}	&0.999515692	&0.99947033\\
$\quad$random\_6	&0.944702392	&\multicolumn{1}{c}{-}	&0.999803805	&0.999046267\\
$\quad$random\_7	&0.944876319	&\multicolumn{1}{c}{-}	&0.999865465	&0.997370819\\
$\quad$random\_8	&0.945166645	&\multicolumn{1}{c}{-}	&0.999905607	&0.998116155\\
$\quad$random\_9	&0.945115357	&\multicolumn{1}{c}{-}	&0.999796598	&0.999199786\\
$\quad$random\_10	&0.945039517	&\multicolumn{1}{c}{-}	&0.999763914	&0.998597517\\\bottomrule
\end{tabular}
\label{tab:EC-bigO-Rvalues}
\end{table}

\subsection{Secondary Findings: Worst Case is Not Worst}
The theoretical worst cases for both Merge Sort~(Figure~\ref{fig:merge_EC_bigO}, pink line) and Bubble Sort~(Figure~\ref{fig:bubble_EC_bigO}, pink line) do not have the worst energy consumption. Instead, their random cases consume the most energy. Bubble Sort random case (Figure~\ref{fig:bubble_EC_bigO}, brown line) is five times slower than Bubble Sort worst case~(pink~line), Merge Sort random case~(Figure~\ref{fig:merge_EC_bigO}, brown line) is two times slower than Merge Sort worst case~(pink~line). 
%

In the following, our results demonstrate that this unexpected \emph{``worst case time complexity case is not the worst case energy''}-phenomena is really a
\emph{``worst case time complexity case is not the worst case wall time''}-phenomena and that the high correlation between energy and wall time carries over to energy consumption. 

\begin{figure*}
    \centering
    \begin{subfigure}[b]{0.33\textwidth}
         \centering
         \includegraphics[width=\textwidth]{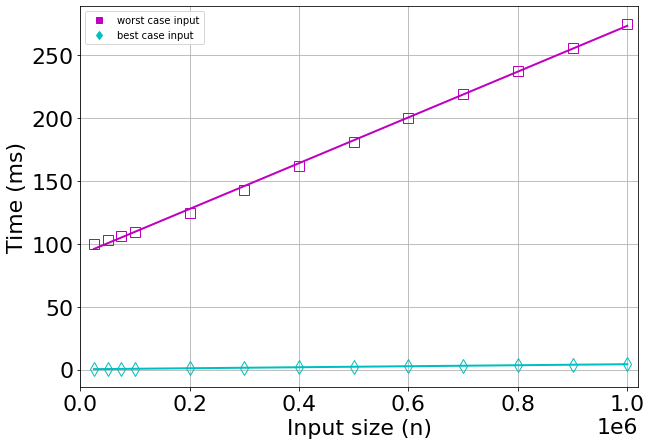}
         \caption{Counting Sort, with time complexity O(n+k)}
         \label{fig:counting-WT-bigO}
     \end{subfigure}
    \begin{subfigure}[b]{0.33\textwidth}
         \centering
         \includegraphics[width=\textwidth]{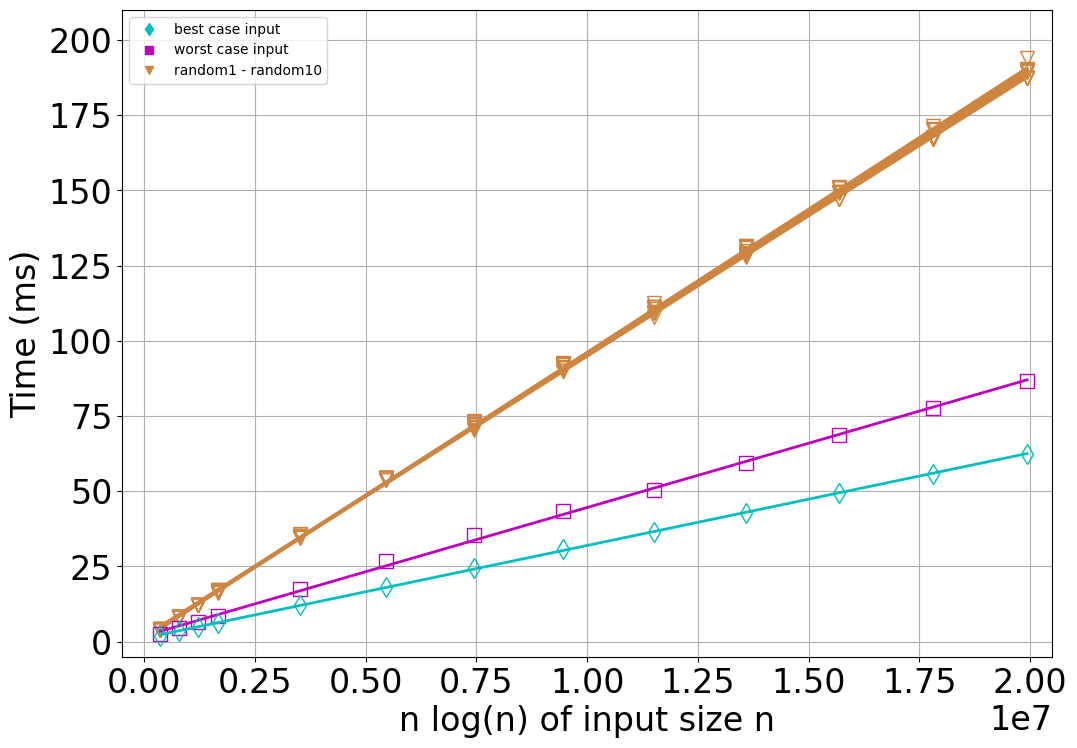}
         \caption{Merge Sort, with time complexity O(n log(n))}
         \label{fig:merge_WT_bigO}
     \end{subfigure}  
     \begin{subfigure}[b]{0.33\textwidth}
         \centering
         \includegraphics[width=\textwidth]{figures/Bubble/bubble_WT_bigO.png}
         \caption{Bubble Sort, with time complexity \bigo($n^2$)}
         \label{fig:bubble_WT_bigO}
     \end{subfigure}  
     \begin{subfigure}[b]{0.33\textwidth}
         \centering
         \includegraphics[width=\textwidth]{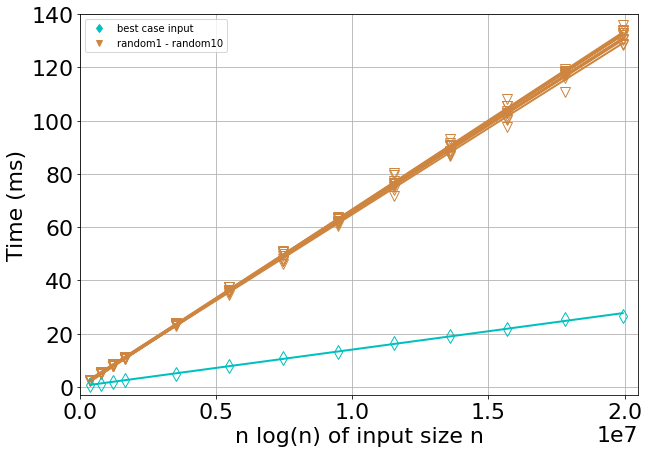}
         \caption{Quick Sort best case and random case, with time complexity O(n log(n))}
         \label{fig:quick1_WT_bigO}
     \end{subfigure}    
     \begin{subfigure}[b]{0.33\textwidth}
         \centering
         \includegraphics[width=\textwidth]{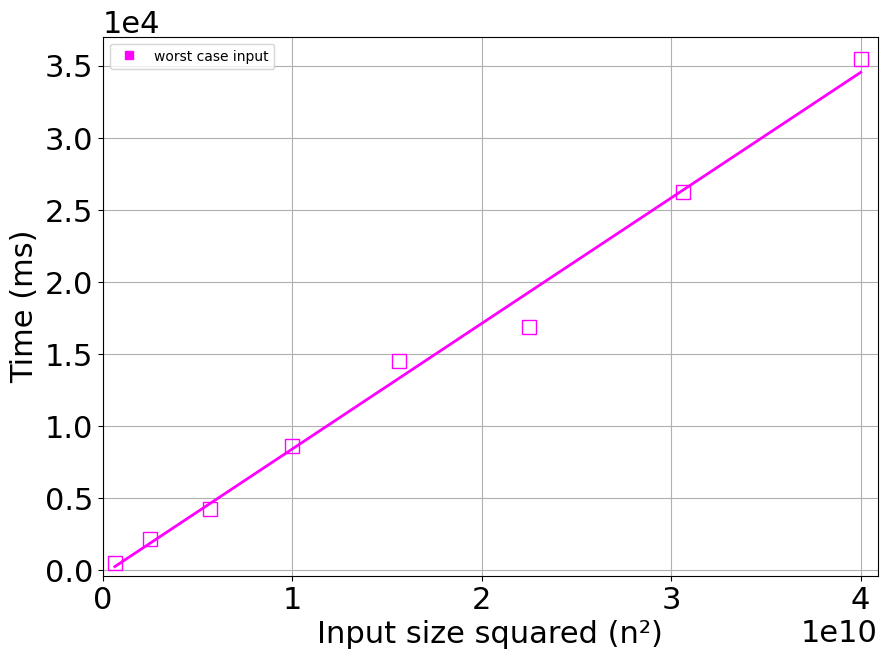}
         \caption{Quick Sort worst case, with time complexity O($n^2$)}
         \label{fig:quick2_WT_bigO}
     \end{subfigure}

    \caption{Wall time against time complexity, where the x-axis is adjusted according to the BigO of the sorting algorithm. Inputs shown are 25000, 50000, 75000, and so on,  where for example if $n = 25 *10^3$ then $n^2 = 6.25 *10^8$. Note the scientific notation on both axes.}
    \label{fig:All-WT-bigO}
\end{figure*}

\paragraph{Time Complexity and Wall Time} \label{sec:sq1}
Figure~\ref{fig:All-WT-bigO} shows the relation between wall time and time complexity for the input space; the input generated to create the theoretical best case are blue, the input generated to demonstrate the theoretical worst case are pink and the random cases are brown. 
More precisely, we depict complexity-converted functions for Bubble Sort~(Figure~\ref{fig:bubble_WT_bigO}), Counting Sort~(Figure~\ref{fig:counting-WT-bigO}), Merge Sort~(Figure~\ref{fig:merge_WT_bigO}) and Quick Sort~(Figures~\ref{fig:quick1_WT_bigO} and~\ref{fig:quick2_WT_bigO}).  

First we observe that similar to the energy measurements, the theoretical worst case inputs for both Merge Sort and Bubble Sort do not yield the worst case wall time for those algorithms, as demonstrated in Figures~\ref{fig:bubble_WT_bigO} and~\ref{fig:merge_WT_bigO}. The random cases (brown lines) have a steeper gradient the worst cases (pink lines). In fact, the Bubble Sort random case (Figure~\ref{fig:bubble_EC_bigO}, brown line) is five times slower than Bubble Sort worst case~(pink~line), Merge Sort random case~(Figure~\ref{fig:merge_EC_bigO}, brown line) is two times slower than Merge Sort worst case~(pink~line). 

The obtained measurements show strong correlations between wall time and various time complexities, with all $R^2$ values $\geq 0.99$, see Table~\ref{tab:WT-bigO-Rvalues}. More than 99\% of the variance of measured wall time is dependant on the linear relationship between the wall time and the converted input size, i.e., the time complexity. 

\begin{table}[]
\centering
\caption{The coefficient of determination ($R^2$) between wall time and time complexity for sorting algorithms $^{4}$.}
\begin{tabular}{@{}l@{\;}l@{\;}l@{\;}l@{\;}l@{}} \toprule
	&Bubble Sort	&Counting Sort	&Merge Sort	&Quick Sort\\ \midrule
Best Case	&0.992769748	&0.996999163	&0.999868557	&0.998162534\\
Worst Case	&0.99999667	&0.998864978	&0.998922612	&0.991910123\\
Random Case	&	&	&	&\\
$\quad$random\_1	&0.999993804	&\multicolumn{1}{c}{-}	&0.999638783	&0.99900458\\
$\quad$random\_2	&0.999993544	&\multicolumn{1}{c}{-}	&0.999782242	&0.999232797\\
$\quad$random\_3	&0.999994293	&\multicolumn{1}{c}{-}	&0.999792173	&0.99945323\\
$\quad$random\_4	&0.9999934	&\multicolumn{1}{c}{-}	&0.999772268	&0.997255538\\
$\quad$random\_5	&0.999995474	&\multicolumn{1}{c}{-}	&0.999370776	&0.999447197\\
$\quad$random\_6	&0.999996593	&\multicolumn{1}{c}{-}	&0.999713262	&0.999087246\\
$\quad$random\_7	&0.999997132	&\multicolumn{1}{c}{-}	&0.999706366	&0.997937353\\
$\quad$random\_8	&0.999997614	&\multicolumn{1}{c}{-}	&0.999835893	&0.998270255\\
$\quad$random\_9	&0.999994853	&\multicolumn{1}{c}{-}	&0.999616117	&0.999274031\\
$\quad$random\_10	&0.99999449	&\multicolumn{1}{c}{-}	&0.99964665	&0.998921056\\ \bottomrule
\end{tabular}
\label{tab:WT-bigO-Rvalues}
\end{table}

\paragraph{Wall time and Energy consumption}
\label{sec:sq2}
The obtained measurements provide evidence of strong correlations between energy consumption and wall time, i.e., $R^2 \geq 0.99$, see Table~\ref{tab:EC_WT-Rvalues}. More than 99\% of the variance of energy consumption can be explained by the linear relationship between the measured energy consumption and wall time. 

\paragraph{Remark} The results demonstrate that the theoretical worst case inputs for Merge Sort and Bubble Sort do not yield the worst case energy consumption and that this can be observed already when considering the simpler metric wall time. The high correlation between wall time and energy consumption, demonstrate that this phenomena is not due to energy being consumed in unexplained ways; it presents itself already in the time consumption. 
 
\begin{table}[]
\centering
\caption{The coefficient of determination ($R^2$) between energy consumption and wall time for sorting algorithms $^{4}$.}
\begin{tabular}{@{}l@{\;}l@{\;}l@{\;}l@{\;}l@{}} \toprule
&Bubble Sort	&Counting Sort	&Merge Sort	&Quick Sort\\ \midrule
Best Case	&0.999997096	&0.998995094	&0.999948762	&0.999898335\\
Worst Case	&0.999998673	&0.995788471	&0.999858448	&0.999898343\\
Random Case	&	&	&	&\\
$\quad$random\_1	&0.999997433	&\multicolumn{1}{c}{-}	&0.999970247	&0.999972279\\
$\quad$random\_2	&0.999997139	&\multicolumn{1}{c}{-}	&0.999986011	&0.999986731\\
$\quad$random\_3	&0.999997472	&\multicolumn{1}{c}{-}	&0.99996755	&0.999970581\\
$\quad$random\_4	&0.999995352	&\multicolumn{1}{c}{-}	&0.99997246	&0.999926295\\
$\quad$random\_5	&0.999997548	&\multicolumn{1}{c}{-}	&0.999956767	&0.999975374\\
$\quad$random\_6	&0.999997604	&\multicolumn{1}{c}{-}	&0.999976442	&0.999954708\\
$\quad$random\_7	&0.99999748	&\multicolumn{1}{c}{-}	&0.999963467	&0.999929902\\
$\quad$random\_8	&0.999997902	&\multicolumn{1}{c}{-}	&0.999982237	&0.999986511\\
$\quad$random\_9	&0.99999743	&\multicolumn{1}{c}{-}	&0.999953085	&0.999963855\\
$\quad$random\_10	&0.999997235	&\multicolumn{1}{c}{-}	&0.99997382	&0.999963179\\\bottomrule
\end{tabular}
\label{tab:EC_WT-Rvalues}
\end{table}

\section{Validity}\label{sec:validity}
In this section, we discuss validity of the study. We identify factors that may interfere with the results of our study (Section~\ref{sec:internalvalidity}), discuss generalization of the results to the concept or theory behind the experiment (Section~\ref{sec:construct-validity}), and discuss generalization of the results to other settings (Section~\ref{sec:external-validity}). 

\subsection{Internal Validity} 
\label{sec:internalvalidity}

\subsubsection{Difference in methodology}
The Java "deep copy" method was introduced to counter the import of input. It produced data with less variance for the more efficient algorithms, allowing trends to be seen. However, this method is not feasible for less efficient algorithms such as Bubble Sort sorting large inputs due to the overflow of the energy register. 

In addition, the ReadCSV energy consumption measurements were taken separately to the sorting algorithms. As the results are based on the ReadCSV data, it would be ideal to improve the methodology such that the true energy consumption of reading the inputs into arrays could be taken into account, or avoid measuring the import of the input altogether, e.g., measuring the energy consumption of only the sorting, i.e., from just before the sorting until after the sorting.

\subsubsection{Not normally distributed data}
When a sample distribution is not normal, we cannot confidently say that the errors that affected the measurements in one run affected the second run in a similar manner~\cite{cruz2021green}. That means that there may be confounding variables, i.e., variables other than the independent variable which could affect the measurements and lead to inaccurate conclusions.
Therefore, we tested to which degree the samples are normal distributed; in Table \ref{tab:ND}, the normal distributions of each algorithm are displayed. Counting Sort was found to be the most normally distributed followed by Merge Sort then Quick Sort, while Bubble Sort shows little to no normal distribution. It is important to note that Quick Sort worst case has the same time complexity as Bubble Sort, O($n^2$). 

\begin{table}[H]
    \centering
    \begin{tabular}{|c|c|c|c|c|}
        \hline
        Algorithm & CPU BC \% & CPU WC \%  & CPU RD \%  \\
        \hline
        Bubble Sort   & 0\%           & 12.5\%          & 58.8\%  \\
        \hline
        Counting Sort & 92.3\%           & 92.3\%           & -  \\
        \hline
        Merge Sort   & 76.9\%           & 84.6\%           & 89.2\%  \\
        \hline
        Quick Sort    & 76.9\%           & 12.5\%          & 87.7\%  \\
        \hline
    \end{tabular}
    \caption{Percentage of normal distribution found in data sets using the Shapiro-Wilk test, per sorting algorithm.}
   
    \label{tab:ND}
\end{table}
Algorithms with time complexity $n^2$, i.e., Bubble Sort and Quick Sort worst case, produced data that was not normally distributed, see Table~\ref{tab:ND}. As mentioned, the approach used for testing linear correlation is independent of whether the distributions are normally distributed and the results remain valid. 

Quick Sort worst case data is 12.5\% normally distributed, similar to Bubble Sort, which indicates that this is not a methodological issue. 
More research is required to find a standard procedure which provides normally distributed data for most algorithms.

\subsubsection{Input ranges}
For best and worst case inputs, the CSVs contained integers ranging from 0 to the length of the input. However, the random case inputs contained integers ranging from -max int to max int. A test revealed that this variance in inputs did not affect the CPU energy consumption. 

\subsubsection{Energy measured}
Alternatively to measuring the CPUs energy consumption, we could have used the energy consumed by a broader set of hardware components, e.g., the RAPL register containing the package energy~\cite{RAPLinAction}. 
In the preliminary study, we found that around 82\% - 86\% of the package energy consumption occurs in the CPU~\cite{BScThesis}. This ratio appears to be steady across the data gathered, therefore it is reasonable to hypothesise that if Package energy consumption is plotted against time complexity, the similar trends would be seen.

\subsubsection{Thermal Throttling}
Considering that there may have been a difference in thermal throttling. Similar to the previous point, the question of whether thermal throttling occurred does not affect the overall results. If throttling did occur but was stable for all tests, then all tests have a greater wall time, meaning they are still comparable to each other.

\subsection{Construct Validity}
\label{sec:construct-validity}

\subsubsection{Algorithm Theory}
The data gathered concurs with textbook theory which states that time complexity can be used to describe run time \cite{introToAlgs}. We see that Bubble Sort, a \bigo($n^2$) algorithm, is the least efficient at sorting inputs, while Counting Sort, a \bigo(n + k) algorithm, is the most efficient as seen in Figure~\ref{fig:All-WT-bigO}. In addition, as stated in standard text, Quick Sort best case, an \bigomega($n \log(n)$) algorithm, is more efficient than Merge Sort best case, also an \bigomega($n log(n))$ algorithm. 

\subsubsection{$E \equiv P \times T$}
Our experiment shows a strong correlation between energy and time as defined in Eq. \ref{eq: energy}. This relationship holds for this study as the experiment was run on single kernels; a similar experiment run on multiple kernels may show a different trend. More research is required to investigate how parallel computing affects the correlation between wall time and energy consumption. 

\subsubsection{Energy and Time Complexity}
The strong correlation between wall time and time complexity across sorting algorithms is evident, as seen in Figure \ref{fig:All-WT-bigO} and supported by Table \ref{tab:WT-bigO-Rvalues}. This similarly strong correlation holds between energy consumption and time complexity across sorting algorithms, shown in Figure \ref{fig:All-EC-bigO} and Table \ref{tab:EC-bigO-Rvalues}, indicating that a new term "energy complexity" may be a valid concept. However, the limitation of this study is that using time complexity as a guideline to predict energy consumption depends on the power draw of the particular algorithm
, and it is not clear what factors affect power draw. More research is required into the power draw of sorting algorithms if a practical guide to choosing algorithms is to be created for software developers.

\subsection{External Validity}
\label{sec:external-validity}
The experiment was carried out on one type of computer, one operating system, one programming language using one type of data structure. 
Changing the hardware may yield different results and there is not yet evidence that allow us to draw similar conclusions for other platforms. 

Choosing an external measurement technique (instead of the software based estimation provided by RAPL) would provide us with the both precise and accurate measures. The experimental design is prepared for measuring externally and can be directly repeated with external measurements. We would expect that a different measurement technique would impact the variance of the obtained measurements since it would measure not only the CPU but the entire system. Thus, while we expect to be able to obtain similar conclusions, we also expect this would require a larger sample size. This aligns with the study by Bunse et al. ~\cite{Bunse2009} which used a sample size of 500 measurements.  

Using a different Java Development Kit may change the relation energy and time complexity. A recent study show that the high correlation between execution time and energy consumption is preserved across JDK 7, 8, 9, 10, 11 and 12~\cite{Kumar2019}. New studies would be required to establish similar relation for all later versions.

Choosing a different programming language would yield a different energy and time consumption~\cite{pereira2021}. However, since time complexity in general holds across programming languages, we expect to obtain similar results. 
Indeed, we expect to obtain similar results using the approach presented in this paper with other programming languages which have an equally low or lower standard deviation in energy consumption, e.g., JavaScript, C++, or C\#, see Table~\ref{tab:proglang}. Because the sample size depends on the standard deviation we would expect to obtain similar results with the same sample sizes, namely 30 (with the exception of Bubble Sort). In case the language executes faster, e.g., C++~\cite{pereira2021}, we may be able to obtain useful results for Bubble Sort using the more precise measurement methodology. However, to establish this, new evidence is required.

\begin{table}
\begin{tabular}{|l|c|c|c|c|}
\hline
\textbf{Language} & \textbf{Median} & \textbf{Std.\ Dev} & \textbf{Average} \\
\hline
Python &  19.79 & 35.72 & 31.21 \\
JavaScript & 0.31 & 1.89 & 1.14 \\
Java &  0.81 & 1.03 & 1.08 \\
TypeScript & 0.17 & 30.94 & 13.99 \\
C++ &  0.24 & 0.91 & 0.73 \\
C\# &  0.51 & 1.20 & 0.99 \\
\hline
\end{tabular}
\caption{The median and standard deviation of the programming languages' energy consumption standard deviations~\cite{pereira2021}. They are ordered according to their average popularity of~\cite{IEEE_Top_Languages_2023,StackOverflow_2023_Survey,Statista_2023_Most_Used_Languages} with most popular first.}
\label{tab:proglang}
\end{table}
%

\section{Related work}\label{sec:related}
In the following we relate the results of this study to the findings in other related studies.
\subsection{Primary Finding: Energy Consumption and Time Complexity}
Mahmud and Hussain study the relation between energy consumption and both time and space complexity of Insertion Sort written in Java~\cite{Mahmud2022}. The experiments are executed on two different architectures and produce a robust result. Compared to our study, they study only worst case time complexity and have chosen the inputs randomly. 

An empirical study by Bunse et al.~\cite{Bunse2009} compares the energy consumption of Bubble Sort, Heap Sort, Insertion Sort, Merge Sort, Quick Sort, Selection Sort, Shaker Sort, and Shell Sort on mobile devices. They vary the input size between 0 and 1000 elements and use external measurement techniques to obtain their energy consumption. They collect 500 measurements for each input size. They test for correlation in a different way. With their methodology, they have not been able to establish correlation between time complexity and energy. One of their main findings is that for mobile environments memory requirements are crucial concerning energy consumption. 

Another study by Schmitt et al.~\cite{Schmitt2021} compares energy-efficiency of six sorting algorithms Merge Sort, Heap Sort, Quick Sort, Insertion Sort, Bubble Sort, and Selection Sort. 
They experiment included two variants of each algorithm implemented in two programming languages, C and Python and they sampled each algorithm on two, out of four, state-of-the-art server systems with different CPUs. Their sample size was five, which is small relative to our setup with 30 and 400. In contrast to our study, their aim was not to correlate the time complexities with energy consumption, but instead understand the practical energy efficiency. 

\subsection{Secondary Findings: Practical Worst Case and Worst Case Time Complexity}
Bubble Sort and Merge Sort random case performs worse than the respective worst cases. A previous study on Merge Sort~\cite{kirkeby2022energy} observed a similar phenomenon on a different type of hardware, an ARM Cortex-A72 processor, showing that this may not be a hardware issue but a software issue. 

In modern computer architectures there are several designs that increase the performance., e.g., pipelines, caches, branch prediction, and out-of-order execution~\cite{Schoeberl2008}.
A preliminary experiment~\cite{PeterSestoft} indicates that this is caused by branch mispredictions; here, random input caused 900 times more branch mispredictions than sorted and reverse sorted inputs.
Further studies are required to establish how the theoretical and the experimental worst case relate.

\section{Conclusion and Future Work}

This investigative study was conducted using experimental methods which allowed us to explore how time complexity, wall time and energy consumption are related. Our results provide empirical evidence that a strong correlation exists between energy consumption and time complexity for four commonly-used sorting algorithms written in Java: Bubble Sort, Counting Sort, Merge Sort and Quick Sort. Three time complexities are examined: $n^2$, $n\log(n)$ and $n + k$. 
High coefficient of determination values found for all four algorithms indicate that energy consumption and time complexity are closely correlated. This correlation is shown especially strong for Merge Sort, Counting Sort and Quick Sort. Given these results, time complexity has the potential to be used as a model for predicting energy consumption of sorting algorithms.


Potential future work includes refining the methodology such that the same approach is suitable for algorithms with different time complexities, as well as extending the experiment to investigate whether a similar strong correlation can be established across varying hardware and different programming languages. 



%

\bibliographystyle{ACM-Reference-Format}
\bibliography{sample-base}


\end{document}